\documentclass[sigconf,screen]{acmart}

\acmConference[FSE '26 Demonstrations]{The 34th ACM Joint European Software Engineering Conference and Symposium on the Foundations of Software Engineering - Demonstrations Track}{December 2026}{City, Country}

\acmBooktitle{FSE '26: ACM Joint European Software Engineering Conference and Symposium on the Foundations of Software Engineering — Demonstrations Track, December 2026, City, Country}
\acmPrice{15.00}
\acmISBN{}
\acmDOI{}

\usepackage{enumitem}
\usepackage{multirow}
\usepackage{mdframed}
\usepackage{framed}

\usepackage{tikz}
\newcommand*\circled[1]{\tikz[baseline=(char.base)]{
            \node[shape=circle,draw,inner sep=2pt] (char) {#1};}}

\usepackage[most]{tcolorbox}
\usepackage{listings}
\usepackage{xcolor}

\definecolor{codebg}{RGB}{248,248,248}
\definecolor{codeframe}{RGB}{180,180,180}
\definecolor{linenum}{RGB}{110,110,110}

\definecolor{codebg}{RGB}{255,255,255}
\definecolor{codeframe}{RGB}{180,180,180}
\definecolor{linenum}{RGB}{110,110,110}

\definecolor{diffRed}{RGB}{215,58,73}
\definecolor{diffGreen}{RGB}{128,128,128}

\lstdefinestyle{paperCode}{
  basicstyle=\ttfamily\footnotesize,
  numbers=left,
  numberstyle=\ttfamily\footnotesize\color{linenum},
  numbersep=3pt,
  xleftmargin=1.5em,      
  frame=none,
  showstringspaces=false,
  tabsize=2,
  breaklines=true,
  breakatwhitespace=true,
  escapeinside={(*@}{@*)}
}

\newtcblisting{codebox}[1][]{
  enhanced,
  listing only,
  listing options={
    style=paperCode,
    #1
  },
  colback=codebg,
  colframe=codeframe,
  boxrule=0.4pt,
  arc=3pt,
  left=0.5pt,
  right=0.5pt,
  top=1pt,
  bottom=1pt
}
\usepackage{textcomp}
  \definecolor{diffstart}{named}{gray}
  \definecolor{diffincl}{HTML}{00a67d} 
  \definecolor{diffrem}{named}{red}

 \lstdefinelanguage{diff}{
    basicstyle=\ttfamily\small,
    morecomment=[f][\color{diffstart}]{@@},
    morecomment=[f][\color{diffincl}]{+\ },
    morecomment=[f][\color{diffrem}]{-\ },
    }

\begin{document}

\title[MOVis: Visualising Missed Patches Across Variants]{MOVis: A Visual Analytics Tool for Surfacing Missed Patches Across Software Variants}

\author{Jorge Gonzalo Delgado Cervantes}
\orcid{0009-0005-5155-8914}
\affiliation{
  \institution{University of Nevada Las Vegas}
  \city{Las Vegas}
  \country{USA}
}
\email{delgaj10@unlv.nevada.edu}

\author{Daniel Ogenrwot}
\orcid{0000-0002-0133-8164}
\affiliation{
  \institution{University of Nevada Las Vegas}
  \city{Las Vegas}
  \country{USA}
}
\email{ogenrwot@unlv.nevada.edu}

\author{John Businge}
\orcid{0000-0003-3206-7085}
\affiliation{
  \institution{University of Nevada Las Vegas}
  \city{Las Vegas}
  \country{USA}
}
\email{john.businge@unlv.edu}

\renewcommand{\shortauthors}{Jorge et al.}

\begin{abstract}
Clone-and-own development produces families of related software variants that evolve
independently. As variants diverge, important fixes applied in one repository are often
missing in others. \texttt{PaReco} has shown that thousands of such \emph{missed opportunity}
(MO) patches exist across real ecosystems, yet its textual output provides limited
support for understanding where and how these fixes should be propagated.
We present \texttt{MOVis}, a lightweight, interactive desktop tool that visualizes MO
patches between a source and target variant. \texttt{MOVis} loads \texttt{PaReco}'s MO
classifications and presents patched and buggy hunks side-by-side, highlighting
corresponding regions and exposing structural differences that hinder reuse. This
design enables developers to quickly locate missed fixes, understand required
adaptations, and more efficiently maintain consistency across software variants. 
The tool, replication package, and demonstration video are
available at \url{https://zenodo.org/records/18356553} and
\url{https://youtu.be/Ac-gjBxHJ3Y}.
\end{abstract}


\maketitle

\section{Introduction}
Clone-and-own development is a widely adopted practice in which developers create
new projects by copying an existing repository and evolving it independently~\cite{Gregorio:2012,Gamalielsson:2014Sustainability,ishio:msr:2017,alexander:ease:2022}.
This practice is pervasive on modern social coding platforms such as GitHub,
which hosts hundreds of millions of repositories and supports large-scale forking
and reuse~\cite{octoverse,Gousios:2014ICSE,Zhou:2020,Businge:chap:2023}. Over time, some forks evolve into
long-lived \emph{software variants} that share a common ancestry but follow
independent development trajectories to add features, address local requirements,
or resolve governance constraints~\cite{Gregorio:2012,Gamalielsson:2014Sustainability,
businge:saner:2022}. As these variants diverge, maintenance becomes fragmented
across teams, increasing the risk of redundant re-implementation, inconsistent
behavior, and unpatched vulnerabilities across the variant family~\cite{businge:emse:2022,Businge:chap:2023}.

A key challenge in variant maintenance is the failure to propagate relevant fixes
from one variant to another. Large-scale empirical studies have shown that long-lived
variants rarely integrate changes across repositories, even when they remain
structurally related~\cite{businge:emse:2022}. Building on this
observation, \texttt{\textbf{PaReco}}~\cite{poedja:fse:2022} is a clone-based analysis
tool that compares patched and buggy code hunks from bug-fix pull requests against
related variants and identifies a substantial number of \emph{missed opportunities}
(MOs): cases where a bug was fixed in a source variant, but the corresponding buggy
code still exists in a related target variant. These missed fixes represent reusable
changes that could reduce maintenance effort and improve reliability, yet remain
unadopted in practice~\cite{ogenrwot2025scam}.

Despite their relevance, MO cases are difficult for developers to act on.
\texttt{PaReco}'s output is purely textual, consisting of patched hunks and the target
files in which their buggy counterparts appear. To understand how a fix should be
propagated, developers must manually locate the corresponding region in the
target repository, search for unpatched lines, and reconstruct how the surrounding
context has drifted over time. \textbf{In practice, this requires developers to mentally
align source and target code across large, independently evolved files using only
raw diffs and file paths.} These challenges are amplified when a pull request
contains multiple hunks across several files, forcing developers to stitch together
fragmented information. As a result, developers lack practical support for
understanding where a missed fix should be applied or what adaptations may be
needed, discouraging patch reuse even when the fix is clearly applicable.

To address this gap, we present \texttt{MOVis}, a lightweight Qt-based tool~\cite{Qt-framework}
that provides an interactive visualisation of missed patches across software variants.
While \texttt{PaReco} can identify missed opportunity (MO) patches at scale, its
output is textual and requires developers to manually locate and align patched and
buggy code across divergent repositories. \texttt{MOVis} overcomes this limitation by
displaying patched source hunks alongside the corresponding still-buggy regions in
the target variant, using side-by-side visualisation and highlighting to make this
relationship explicit. This externalises the source--target alignment that
\texttt{PaReco} leaves implicit, enabling developers to act on missed opportunities
with minimal manual effort.

\texttt{MOVis} loads \texttt{PaReco}’s MO classifications produced by \texttt{PaReco} and displays the
patched source hunk alongside the relevant target code with syntactic alignment and
highlighting of missing changes. As developers select a patched hunk in the source
view, the corresponding buggy region in the target variant is automatically
located and highlighted. This eliminates the need to manually search large,
drifted files or mentally reconstruct how a fix should be propagated, allowing
developers to quickly see \emph{where} a missed fix should be applied and
\emph{how} the surrounding context differs. \texttt{MOVis} is an interactive,
developer-facing tool designed for exploratory inspection of real-world missed
patches, making it well suited for demonstration in realistic
variant-maintenance scenarios.

\subsection{Motivating Example}

To illustrate the practical challenges of acting on missed opportunity (MO) reports
produced by \texttt{PaReco}, we present a concrete MO drawn from the \texttt{PaReco}
dataset. The source variant in this example is the official \texttt{apache/kafka}~\cite{apache-kafka}
repository, and the target variant is the forked \texttt{linkedin/kafka}~\cite{linked-kafka} repository,
which was originally derived from the upstream Apache Kafka codebase and has since
evolved independently, both hosted on GitHub. According to \texttt{PaReco}, the two
variants diverged on June~2,~2022. The MO arises from pull request (PR)~\#12842 in the
source variant,\footnote{\url{https://github.com/apache/kafka/pull/12842/files}}
which modifies the file
\path{streams/src/main/java/org/apache/kafka/streams/state/internals/RocksDBStore.java}.
We focus on the first affected hunk of this file, which is representative of the
localization challenges that also arise in larger, multi-hunk pull requests.

\begin{lstlisting}[xleftmargin=3em,framexleftmargin=3em, language = diff, caption = Source (patched hunk) PR--12842 from the source variant, label={listing:source_hunk}, frame = single,basicstyle=\ttfamily\scriptsize, breaklines=true, numbers=left, stepnumber=1, xleftmargin=.5cm, firstnumber=584,
xrightmargin=.1cm]
@@ -584,9 +584, 6 @@
void flush() throws RocksDBException;

- void prepareBatchForRestore(final Collection<KeyValue<byte[], byte[]> > records,
- final WriteBatch batch) throws RocksDBException;
void addToBatch(final byte[] key,
    final byte[] value,
    final WriteBatch batch) throws RocksDBException;
\end{lstlisting}

\begin{lstlisting}[xleftmargin=3em,framexleftmargin=3em, language = diff, caption = Unpatched lines (red text) in file \texttt{RocksDBStore.java} in the git\_head--\texttt{fdb9fd0} of target variant, label={listing:target_code}, frame = single, basicstyle=\ttfamily\scriptsize, breaklines=true, numbers=left, stepnumber=1, xleftmargin=.5cm, firstnumber=541, xrightmargin=.1cm, escapechar=!]
long approximateNumEntries() throws RocksDBException;

void flush() throws RocksDBException;

!\textcolor{diffrem}{void prepareBatchForRestore(final Collection<KeyValue<byte[], byte[]> > records,}!
!\textcolor{diffrem}{final WriteBatch batch) throws RocksDBException;}!

void addToBatch(final byte[] key,
        final byte[] value,
        final WriteBatch batch) throws RocksDBException;
\end{lstlisting}

Listing~\ref{listing:source_hunk} shows the patched hunk in the source variant after
PR~\#12842 was merged. In this hunk, the method
\texttt{prepareBatchFo\-rRestore} is removed from the \texttt{RocksDBStore} class.
Listing~\ref{listing:target_code} shows the corresponding code region in the target
variant at commit \texttt{fdb9fd0},\footnote{\url{https://github.com/linkedin/kafka/blame/3.0-li/streams/src/main/java/org/apache/kafka/streams/state/internals/RocksDBStore.java}}
where the same method is still present. The removed lines in the source and the
retained lines in the target are highlighted consistently in red to emphasize that
the target variant still contains code that has been eliminated upstream.

This change constitutes a missed opportunity because PR~\#12842 removes an API
method that upstream developers eliminated to simplify the
\texttt{RocksDBStore} interface and its restore logic, as described in the PR
description and commit message.\footnote{\url{https://github.com/apache/kafka/pull/12842}}
Retaining this method in the target variant may increase maintenance overhead,
complicate future software evolution, and potentially mislead developers into
relying on behavior that is no longer exercised in the upstream codebase. Based
on inspection of the surrounding interface and restore logic, the change appears
applicable to the target variant without additional adaptation.
\texttt{PaReco} therefore classifies this case as an MO because the target variant
could have adopted the upstream simplification but did not.

This example highlights a core challenge that MOVis addresses: although the source
and target hunks correspond to the same method, independent evolution has shifted
their line numbers and surrounding context. While \texttt{PaReco} reports the source
hunk and target file path, it does not reveal the exact location of the buggy code
or provide side-by-side context, forcing developers to manually search and
reconstruct the alignment.

\section{Tool Overview}
\texttt{MOVis} is designed to help developers inspect missed opportunity (MO) patches identified by \texttt{PaReco}. The tool provides an integrated interface in which users can browse MO pull requests, select individual hunks, and compare patched and buggy regions side-by-side.

\subsection{Workflow}
Given \texttt{PaReco}'s MO classification file and the files of the source $\rightarrow$ target repositories, MOVis:

\begin{enumerate}[leftmargin=*]
    \item loads all MO pull requests and their associated hunks;
    \item retrieves source and target file snapshots via the GitHub API using commit and file-path metadata produced by \texttt{PaReco};
    \item locates the corresponding source and target code regions for a selected hunk;
    \item renders the patched (source) and buggy (target) hunks side-by-side and,
          as developers click within the source hunk, automatically highlights the
          corresponding lines in the target variant with synchronized navigation.
\end{enumerate}

Although MOVis is presented in this paper for inspecting Missed Opportunities (MOs), its underlying architecture and interface are generic, and can also visualize other PaReco classifications (e.g., Effort Duplication) using the same source $\rightarrow$ target alignment and highlighting mechanisms.

\vspace{-6pt}
\begin{figure*}[t]
    \centering
    \includegraphics[width=\linewidth]{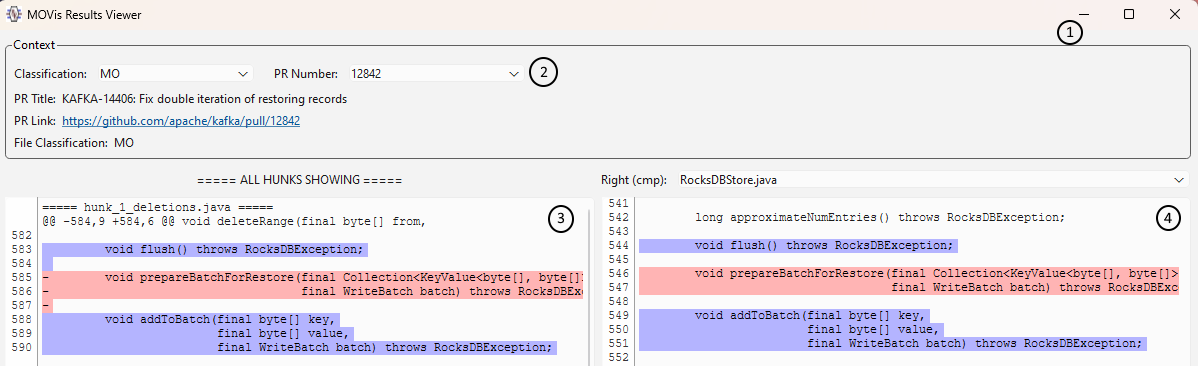}
    \caption{MOVis interface showing the metadata panel (top), Hunks Viewer
    (lower-left), and target variant file (lower-right).}
    \label{fig:interface}
\end{figure*}

\subsection{Interface Components}

The interface is organized into four coordinated regions
(cf. Figure~\ref{fig:interface}):

\begin{itemize}[leftmargin=0.2in]
    \item[\circled{1}] \textbf{Top Panel Group Box.}
    Contains a combo box that allows developers to select a classified pull request produced by \texttt{PaReco}.

    \item[\circled{2}] \textbf{Top Panel Metadata.}
    Displays key information about the selected file, including file classification, pull request title, and a link to the original pull request. The panel updates automatically as users navigate among files.

     \item[\circled{3}] \textbf{Hunks Viewer (lower-left).}
    Displays all hunks associated with a selected file in the current MO pull request.
    When a file is selected (cf.~\circled{4}), this view aggregates every hunk from that
    file, allowing developers to inspect multiple related changes together rather than
    one hunk at a time. Entries such as \texttt{hunk\_[n]\_deletions.java} and
    \path{hunk_[n]_additions.java} represent the removed and added lines of the same
    patch hunk, respectively. Contextual lines are highlighted in blue, added lines in green, and removed lines
    in red, enabling developers to quickly distinguish unchanged context from upstream
    fixes and deletions. This file-level aggregation is particularly useful for
    understanding multi-hunk patches and assessing whether all related changes should
    be propagated to the target variant.

     \item[\circled{4}] \textbf{Target Viewer/Browser (lower-right).}
    Displays (i) a combo box to select a file modified by the pull request and
    (ii) the contents of the selected file from the target variant. Lines that
    correspond to the selected source hunk are highlighted in matching colors,
    enabling direct visual comparison with the Hunks Viewer~\circled{3}.

\end{itemize}

\noindent Figure~\ref{fig:interface} illustrates \texttt{MOVis} using the motivating example from Section~1.1. The left pane (Hunk Viewer)~\circled{3} shows the patched source hunk from the \texttt{apache/kafka} variant, where the obsolete method has been removed upstream, while the right pane (Target Viewer)~\circled{4} displays the corresponding file from the \texttt{linkedin/kafka} variant, in which the same code is still present. Highlighted lines indicate code that was deleted in the source but remains in the target, making the missed opportunity explicit despite differences in line numbers and surrounding context. By visually aligning these semantically corresponding regions, \texttt{MOVis} resolves the localization difficulty illustrated in Listings~\ref{listing:source_hunk} and~\ref{listing:target_code}, allowing developers to immediately identify where an upstream fix applies within a large and independently evolved target file.

Although the motivating example focuses on a deletion-only MO, \texttt{MOVis}
supports the full range of MO scenarios identified by \texttt{PaReco}, including
(i) deletion-only fixes, (ii) fixes that both remove and add code, and
(iii) addition-only fixes where missing logic must be introduced in the target
variant. In all cases, MOVis uses the same source–target alignment and highlighting
mechanism to localize the affected regions and expose how the fix was applied
upstream.


\section{Architecture and Implementation}
\label{sec:architecture}
\texttt{MOVis} is implemented as a standalone desktop application in C++ using the Qt
framework~\cite{Qt-framework}. Its architecture is organised into four layers that support
optional execution of \texttt{PaReco}, parsing of \texttt{PaReco}'s output, localization of
corresponding source and target code, and side-by-side visualization of MO
hunks.

\subsection{Overall Architecture}

\begin{figure}[t]
    \centering
    \includegraphics[width=\linewidth]{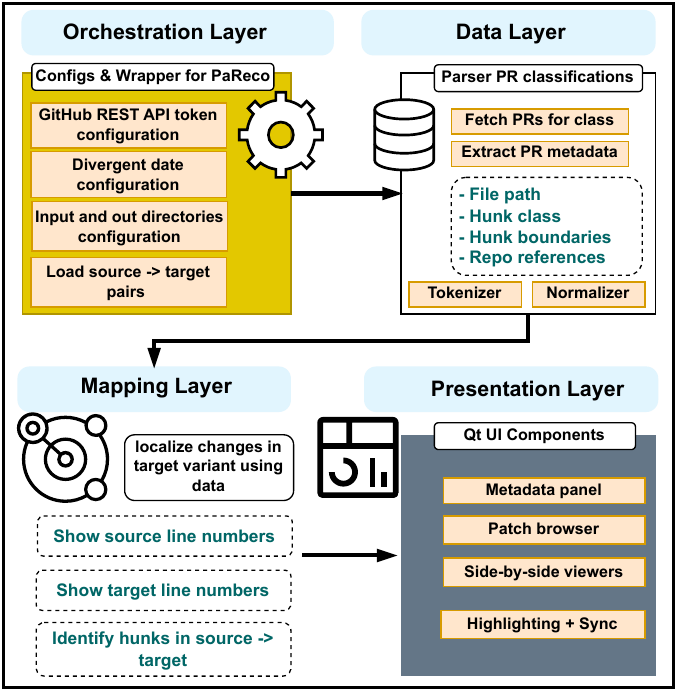}
    \caption{High-level architecture of \texttt{MOVis}.}
    \label{fig:architecture}
\end{figure}

Figure~\ref{fig:architecture} summarizes how these layers interact during execution. The design enables \texttt{MOVis} to either execute \texttt{PaReco} directly or operate independently on previously generated \texttt{PaReco} outputs.

\begin{itemize}[leftmargin=*]

    \item \textbf{Orchestration Layer.}
    Acts as a configuration and control layer that coordinates the execution
    of \texttt{PaReco} and the loading of its results. Through this layer, users may
    optionally run the \texttt{PaReco} command-line tool from within \texttt{MOVis} by providing required parameters such as GitHub access tokens, divergence dates, variant repository identifiers, and output locations. Alternatively, the orchestration layer allows users to bypass \texttt{PaReco} execution and directly load previously generated MO classifications for inspection.
    
    \item \textbf{Data Layer.}
    Responsible for parsing and managing \texttt{PaReco}'s classification output.
    Upon selection of a MO pull request, this layer
    extracts all relevant metadata, including (i) affected file paths,
    (ii) hunk classifications, (iii) hunk boundaries, and (iv) source and
    target repository references. The Data Layer provides a structured,
    queryable representation of MO cases to downstream components. 

    \item \textbf{Mapping Layer.}
    Uses metadata provided by the Data Layer to localize corresponding code
    regions across source and target variants. Given hunk boundaries and file
    references, this layer aligns patched source hunks with their buggy
    counterparts in the target variant, computing accurate line-number
    mappings for both versions. This explicit source $\rightarrow$ target localization
    decouples alignment logic from presentation concerns and supports robust
    navigation across structurally divergent files.
    
    \item \textbf{Presentation Layer.} Implements the Qt-based user interface, including the metadata panel,
    hunk viewer, and side-by-side code viewers. This layer renders aligned
    source and target code regions, manages synchronized scrolling and
    navigation, and applies consistent visual highlighting to emphasize
    added, removed, and unchanged (context) lines across variants.
    
\end{itemize}

\vspace{-6pt}
\vspace{-6pt}

\subsection{Implementation Details}
The following subsections briefly describe how each architectural layer is realized in the current prototype. Qt serves as the backbone of \texttt{MOVis}, providing the
infrastructure required to implement and integrate these layers.

\subsubsection{Orchestration Layer}
The orchestration layer is implemented using Qt's form editor, enabling rapid
construction of the graphical interface while maintaining a clear separation
between configuration and execution logic. User-interface components such as
\texttt{QPushButton}s, \texttt{QPlainTextEdit}s, \texttt{QDateTimeEdit}s, and
\texttt{QLineEdit}s are connected to a command-line version of \texttt{PaReco} and are
used to configure and optionally execute \texttt{PaReco} directly from within \texttt{MOVis}.

To support exploratory analysis without repeated computation, \texttt{MOVis} also allows users to inspect previously generated \texttt{PaReco} outputs via a \texttt{``View Previous Results''} option. This bypasses \texttt{PaReco} execution and
transfers control directly to the \textbf{Data Layer}.

\subsubsection{Data and Mapping Layer}
The data and mapping layers operate dynamically, retrieving information directly from \texttt{PaReco}'s output as required. After a classification and corresponding pull request are selected, the \textbf{Data Layer} extracts all relevant metadata,
including file paths, commit identifiers, and hunk boundaries. The
\textbf{Mapping Layer} then uses this metadata to localize corresponding source and target code regions, which are propagated to the \textbf{Hunk and Target Browsers} for visualization.

These browsers are implemented using custom classes built on top of Qt's widget framework, enabling fine-grained control over rendering and user interaction.

\subsubsection{Presentation Layer}
The \textbf{Hunk and Target Browsers} are implemented using two custom Qt-based
classes:

\begin{itemize}[leftmargin=*]
    \item \textbf{LineNumbers} renders accurate line numbers for both the selected hunk and the target file. It computes line offsets based on the changes reported in the pull request and uses \texttt{QPainter} to draw line numbers within a dedicated \texttt{QWidget}.
    \item \textbf{SimilarityInterface} renders the code content and coordinates updates with the \texttt{LineNumbers} component to maintain consistent alignment during navigation. It primarily leverages
    \texttt{QPlainTextEdit} and \texttt{QWidget} to display and update patched and buggy code regions.
\end{itemize}

The \texttt{SimilarityInterface} class also manages the \texttt{QComboBox}
widgets used to select file paths for both hunk and target views, enabling
efficient context switching when analyzing pull requests that span multiple
files or hunks.

\subsection{Deployment}
\texttt{MOVis} is built and deployed using Qt's \texttt{qmake} build system, enabling straightforward
cross-platform compilation. With appropriate configuration in the project's
\texttt{.pro} file, \texttt{MOVis} can be built for Windows, x86\_64 Linux, and ARM-based
Apple operating systems. To further simplify setup and ensure environment consistency, \texttt{MOVis} can also be deployed using Docker. The Docker configuration encapsulates all build-time and runtime dependencies, allowing users to build and run \texttt{MOVis} without manual toolchain installation.

\section{Related Work}

\texttt{MOVis} builds directly on \texttt{PaReco}~\cite{poedja:fse:2022}, which detects
missed opportunity (MO) patches by analyzing patch reuse across software variants at
scale. While \texttt{PaReco} effectively identifies reusable changes, it presents MO
cases as textual artifacts, leaving developers to manually locate and interpret the
corresponding buggy regions in target variants. \texttt{MOVis} complements
\texttt{PaReco} by providing an interactive, developer-facing visualization that
supports inspection and localization of missed fixes.

More broadly, prior work on software variants has shown that clone-and-own systems
frequently evolve in isolation, with limited coordination and reuse of changes across
repositories. Empirical studies across ecosystems such as Android, npm, and
blockchain platforms report widespread divergence and redundant maintenance in
long-lived variants~\cite{businge:2018icsme,Businge:benevol:2020,businge:blockchain:2022,businge:emse:2022}.
Qualitative studies further indicate that socio-technical factors, such as limited
awareness of relevant changes and difficulty assessing integration risk, contribute
to missed reuse in practice~\cite{businge:saner:2022}. \texttt{MOVis} addresses these
challenges by focusing on developer-centered support for understanding and acting on
missed patches, rather than automating reuse decisions.

\section{Conclusion}

\texttt{MOVis} offers an interactive, developer-focused interface for exploring missed
opportunity patches across software variants. By bridging \texttt{PaReco}'s backend
analysis with intuitive, side-by-side visualization of source and target code,
\texttt{MOVis} reduces the cognitive effort required to understand divergence and localize
missed fixes. The tool supports practical inspection of real-world variant pairs
and is well suited for demonstrating how visualization can aid patch reuse in
clone-and-own development settings.


\bibliographystyle{ACM-Reference-Format}
\bibliography{references}

\end{document}